%% file: main.tex
\documentclass[sigconf]{acmart}
\AtBeginDocument{%
  }

\setcopyright{acmlicensed}
\copyrightyear{2018}
\acmYear{2018}
\acmDOI{XXXXXXX.XXXXXXX}
\acmConference[Conference acronym 'XX]{Make sure to enter the correct
  conference title from your rights confirmation email}{June 03--05,
  2018}{Woodstock, NY}
\acmISBN{978-1-4503-XXXX-X/2018/06}

\usepackage{multirow} 
\usepackage{xspace} 
\usepackage{xcolor} 
\usepackage{graphicx} 
\usepackage{subcaption} 
\usepackage{xcolor, colortbl}
\usepackage[dvipsnames]{xcolor}
\usepackage{enumitem} 
\usepackage{dirtytalk} 
\usepackage{listings}
\usepackage{booktabs}
\usepackage{tabularx}
\usepackage{float}

\newcommand{\figlabel}[1]{%
\begingroup
\setlength{\fboxsep}{1pt}%
\fcolorbox{black}{black}{\textcolor{white}{\scriptsize #1}}%
\endgroup
}

\setcopyright{none}

\begin{document}

\title[Elicitive User Interfaces]{Elicitive User Interfaces: Designing How Users Shape Generative~Interfaces}

\author{Eunhye Kim}
\authornote{Both authors contributed equally to this research.}
\affiliation{%
  \institution{School of Computing, KAIST}
  \city{Daejeon}
  \country{Republic of Korea}
}
\email{gracekim027@kaist.ac.kr}

\author{Bryan Min}
\authornotemark[1]
\affiliation{%
  \institution{University of California San Diego}
  \city{La Jolla}
  \state{California}
  \country{USA}
}
\email{bdmin@ucsd.edu}

\author{Haijun Xia}
\affiliation{%
  \institution{University of California San Diego}
  \city{La Jolla}
  \state{California}
  \country{USA}
}
\email{haijunxia@ucsd.edu}

\author{Juho Kim}
\affiliation{%
  \institution{School of Computing, KAIST}
  \city{Daejeon}
  \country{Republic of Korea}
}
\email{juhokim@kaist.ac.kr}

\renewcommand{\shortauthors}{Eunhye Kim*, Bryan Min*, Haijun Xia, and Juho Kim}

\newcommand{\sysname}{\textsc{Ditto}}

\begin{abstract}

\input{sections/00_abstract}

\end{abstract}


  
\begin{teaserfigure}
\input{sections/00_teaser}
\end{teaserfigure}

\received{20 February 2007}
\received[revised]{12 March 2009}
\received[accepted]{5 June 2009}

\maketitle

\input{sections/01_introduction}
\input{sections/02_relatedwork}
\input{sections/04_designspace}

\input{sections/05_probe}
\input{sections/06_userstudy}

\input{sections/07_longstudy}

\input{sections/08_discussion}

\input{sections/09_conclusion}

\bibliographystyle{ACM-Reference-Format}
\bibliography{cleaned_references}

\input{sections/99_appendix}


\end{document}

%% file: sections/00_abstract.tex
Generative user interfaces (GenUI) promise personalized interfaces to a user's tasks and needs. However, user needs are often implicit---difficult for systems to infer and users to articulate, making it hard for users to arrive at their ideal interface. We propose \textbf{\textit{Elicitive User Interfaces}}, a design approach to GenUI that generates elicitation techniques as part of the interface itself. Elicitive UIs adapt these techniques to the user, task, and interface to draw out user preferences. To guide the design of Elicitive UIs, we synthesize a six-axis design space that shapes how an interface elicits user preferences.
Across two user studies with a design probe, we found that while Elicitive UIs surfaced preferences users had not already formed, and that responses to elicitation varied more across users than across tasks. Users developed more consistent preferences for how they wanted to be elicited, suggesting an opportunity to personalize elicitation itself.

%% file: sections/00_teaser.tex
\centering
 \includegraphics[width=0.9\textwidth]{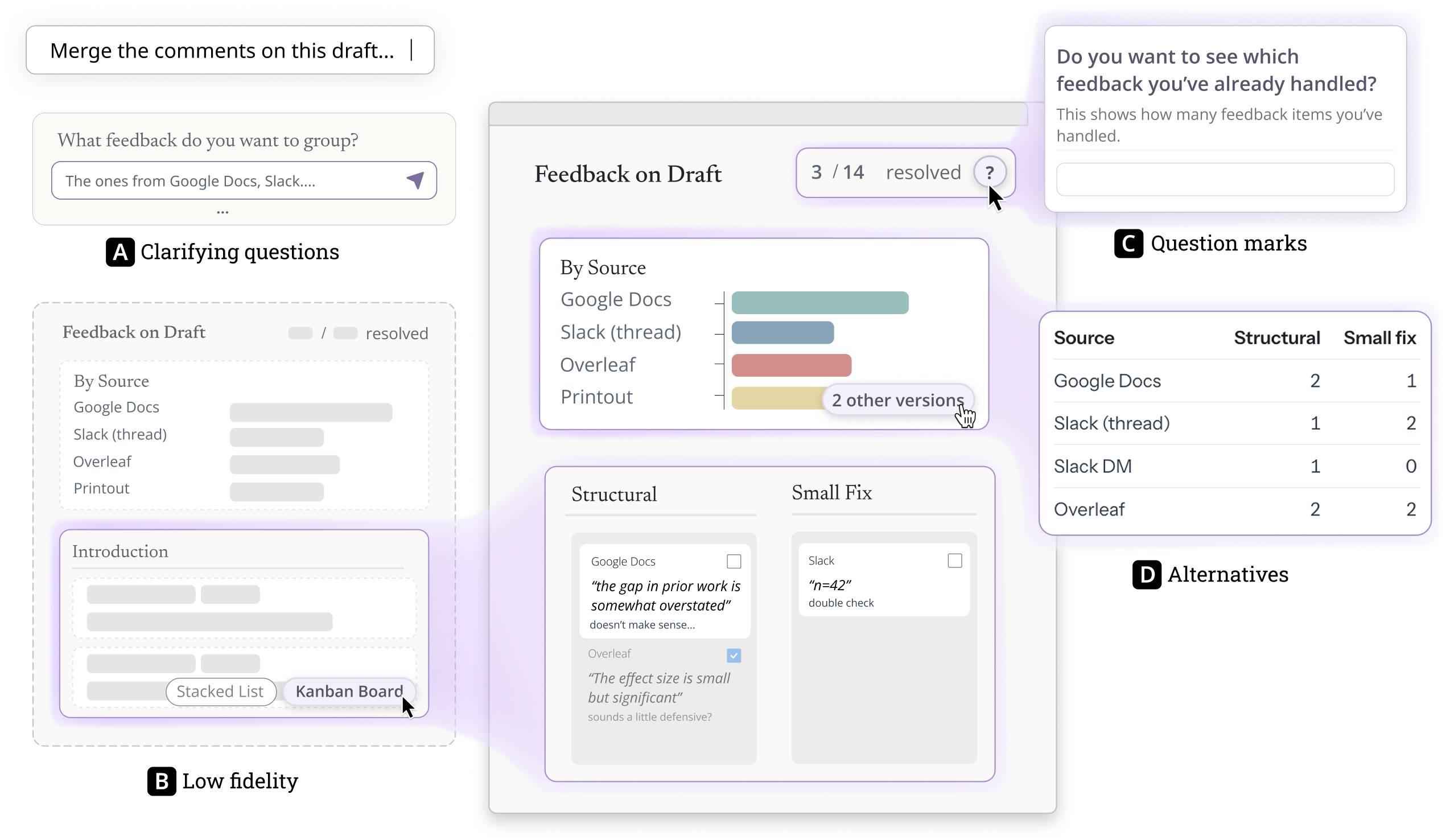}
  \caption{Elicitation techniques on a generated UI. Asked to merge co-author
feedback scattered across sources, the system first (\figlabel{A}) asks the user what feedback to gather. Then, the system generates (\figlabel{B}) a low fidelity sketch and offers candidate forms. The user
switches the stacked list to a kanban board and renders it.
 The rendered UI shows a (\figlabel{C}) question mark and (\figlabel{D}) an alternatives mark, through which the user can change the form or request a change. The user can form and express preferences through EUIs, tailoring the interface to their needs.}
 \Description{A generated interface shown at two stages. A chat prompt across the
top reads, ``Merge the comments on Google Docs, Slack, Overleaf on this draft.''
The left panel, labeled ``Low fidelity version,'' shows a grey sketch of a
``Feedback on Draft'' interface. It includes placeholder counts, a ``By Source''
bar chart for Google Docs, Slack (thread), Overleaf, and Printout, source tabs,
and skeleton comment cards under an ``Introduction'' section. At the bottom,
the user can choose between ``Stacked List'' and ``Kanban Board'' before clicking
``render it?''; the cursor is on ``Kanban Board.'' A purple band connects this
sketch to the rendered interface in the center. The rendered version shows
``3 / 14 resolved,'' a completed ``By Source'' bar chart, source tabs, and a
two-column board labeled ``Structural'' and ``Small Fix'' containing comment
cards. One card is checked off. A question-mark icon next to the resolved count
opens a popover asking whether the user wants to see which feedback they have
already handled and whether this matches how they work. A ``2 other versions''
control below the chart opens a table showing an alternative breakdown of
feedback by source, with counts for structural comments, small fixes, and totals.}
  \label{fig:teaser}

%% file: sections/01_introduction.tex
\section{Introduction} 

The north star of generative user interfaces (GenUIs) is to provide users with personalized interfaces tailored to any task they take on. As coding agents and generative AI models have made it possible to now generate complete interfaces with ease, people are increasingly adopting GenUI into their workflows through chatbots \cite{claude2024artifacts, googleDiscoLabs, openai2025chatgptapps} and no-code development platforms \cite{lovable2026, claude2024artifacts}, creating personal interfaces to plan trips, visualize new concepts, manage research data, create video games, and more.

An effective generative interface hinges on understanding what the user's goals and needs are. However, these needs are often unarticulated and unformed.
Although research works to build better contextual understanding of the user's needs \cite{shaikh2025gum, kim2026maru, alves2026interactiondata, jelly, lam2026jit}, there are inevitably gaps that can be critical to the user. At the same time, generating an interface requires the system to make many decisions the user never would have requested.

For example, a researcher might ask for an interface that unifies co-authors' feedback scattered across Google Docs, Slack, Overleaf and photos of a marked-up printout (Figure~\ref{fig:teaser}). A GenUI system might generate a list of comments, treating every comment as equally important, but the researcher may instead prefer an interface that groups comments by significance, surfacing the key points of feedback. This preference may not exist in a form the researcher can articulate beforehand---the finished list itself can make the organization appear settled, while offering little indication that another form is possible. The generated interface therefore does more than reflect what the system knows about the user, it also shapes whether the users can recognize and express what they might want differently.

To draw out more context that the system cannot infer, GenUI systems additionally elicit user needs by prompting users to respond by choosing a set of answers or typing back to the system \cite{linden1997interactive, li2023elicitinghumanpreferenceslanguage}.
However, user interfaces consist of much more than just text.
Users may have needs about different layouts, styles, media, visual representations, or interaction behavior, yet the textual medium that prompting imposes makes these needs difficult to convey.
Instead, we need to provide broader techniques for eliciting user needs for adapting generative interfaces.

We argue that \textbf{GenUI requires designing not just the interface but also the interactions through which users discover and express what they want from it}.
To do so, we conceptualize a design space of these interactions, which elicit user needs by exposing decisions about how the interface could change, that users might not otherwise think to make.
We distinguish this class of generative interfaces as \textbf{\textit{Elicitive User Interfaces}} (EUIs). EUIs are interfaces that bake in elicitation techniques directly into the design, choosing the techniques best suited to the user's task, their preferences, and the interface they generate. This surfaces more opportunities for the user to improve the interface, helping them arrive at better tailored interfaces to their task and preferences.

Following the previous example of a researcher gathering feedback, a generated interface may first prompt the user with clarifying questions about what pieces of feedback they want to group, as text prompting still has its merits for describing textual elements \cite{linden1997interactive, li2023elicitinghumanpreferenceslanguage}.
But if they cannot express the different groupings into words, generating \textit{multiple alternatives} can complement this, helping them recognize the group they need \cite{gajos2005arnauld}.
Choosing how to lay out those groups may instead benefit from generating the interface's layout in a lower \textit{fidelity}, a technique that invites the researcher to imagine layouts beyond the options the model could suggest \cite{buxton2007sketchingUX, gao, min2026gradual}, prompting them to envision a kanban board over a linear list.
Furthermore, the same decision might be elicited with different levels of \textit{salience} and \textit{assertiveness} by folding the clarifying questions into on-hover tooltips throughout the interface, letting the researcher specify their needs on demand.

GenUI interfaces can leverage this design space to provide the appropriate elicitation techniques according to the user and their task, effectively adapting not only the interface, but the elicitation technique to the user as well.

We investigate how elicitation can be designed as a part of a generative interface, and how these design choices shape users' experiences and responses.

First, we developed a design space of elicitation techniques for a generative interface to adopt.
From a formative study (N=10) observing what modifications users make when elicited and an iterative design process incorporating the literature of elicitation techniques across HCI systems, we derived a design space of EUIs spanning six axes: fidelity, salience, amount, frequency, assertiveness, and placement.
These design space axes characterize how elicitation is presented within the generated interface, including how settled the interface appears, how noticeable or assertive an elicitation is, how much is elicited, and where and when it appears.

To better understand how EUIs help surface user preferences and shape interaction, we conducted two user studies with a design probe that generates EUIs drawn from our design space.
In the first study, we observed that users (N=12) found their resulting interfaces and the generated elicitation methods useful for their task.

We additionally observed that responses to elicitation varied more across users than across tasks, suggesting that elicitation may benefit from being designed around who the user is, not only what they are doing. This finding surfaced our curiosity in whether EUIs are also well suited to become consistent profiles that users can apply across their GenUI tasks, and motivated us to conduct a three-day deployment study (N=3) over repeated use of generated EUIs.
After three days, participants' elicitation preferences crystallized into more salient preferences, but we also observed that EUIs helped some participants reflect on their own preferences and further tune them to steer UI generation.
These findings reinforce our perspective that generating EUIs to elicit user needs demonstrates a promising avenue toward designing generative interfaces for end-users.

In summary, we contribute the following:
\begin{enumerate}
    \item The conceptualization of \textit{Elicitive User Interfaces}, a design approach for generative user interfaces in which elicitation techniques are generated as part of the interface itself, adapted to the user's task, preferences, and the interface it inhabits.
    \item A design space of elicitation techniques that GenUI systems can draw on to generate Elicitive UIs across diverse user tasks and preferences.
    \item Probe studies suggesting that generated EUIs can effectively elicit user preferences, but are more effective when adapted to user preferences over task needs.
\end{enumerate}

This paper instantiates our vision of an emerging need to not only generate interfaces, but to also \textit{design} how they are generated to best leverage their adaptive properties.
We hope our work contributes insights toward this broader goal, helping designers and developers build GenUI systems that better align to users.

%% file: sections/02_relatedwork.tex
\section{Related Work} 

\subsection{Generating Personalized User Interfaces with Generative AI}

The goal of personalized user interfaces has been to provide users with software interfaces that best suit their activities, goals, and environments \cite{fan2006personalizationperspectives, jelly}. If users can get a more personalized interface, they spend less time context-switching \cite{chang2021tabs} or foraging for information \cite{pirolli1999infoforaging} and more time on the task itself, while giving them a greater sense of control over their tools \cite{alves2026interactiondata, marathe2011drivescustomization, mackay1991triggers}. As generative AI models and coding agents have become capable of automatically creating nearly any interface from a user prompt, generative UI as a design approach has grown increasingly popular as a way to give users a personalized user interface.

As we have seen GenUI increase in capabilities, HCI has explored where we can generate interfaces to supplement existing interfaces for a user goal.
For example, after an AI edits an artifact from a user's prompt, the interface can generate an interactive widget, such as a slider or picker mapped to the changed parameters, that lets the user refine the result through direct manipulation instead of prompting again \cite{priyan2024dynavis, vaithilingam2019bespoke, leung2025squire}.
GenUI can also help users further steer interactions with generative AI in conversations \cite{drosos2025promptmiddleware}, code editors \cite{cheng2024biscuit}, command-line interfaces \cite{kasibatla2025command}, and web browsing \cite{kim2026insitu, jiang2026orca}.

In contrast, another line of work instead generates interfaces that primarily serve the user's goal, focusing on tools to supplement the generative interfaces---effectively building a ``tech stack for GenUI'' \cite{kim2026maru, lindley2026genui}.
These works primarily aim to build tools, layers, and intermediate representations that supplement GenUI systems with richer context about users to generate more aligned interfaces. These tools have come in the form of building and maintaining a structured model of the user's task \cite{jelly, lam2026jit} where the goal is to maintain an up-to-date understanding about the user and their activities to generate an interface that can most appropriately support their activity at any moment. Tools have also come in the form of layers---intermediate stages in a generation pipeline---instructing GenUI systems to generate specifications such as state machine diagrams to enrich the interface's interaction behavior \cite{chen2025generative} or information architecture rules that explicitize which elements are important enough to persist in the interface \cite{kim2026maru}.

Our work contributes to the broader ecosystem of GenUI tools and methods by contributing design approaches for GenUI. Specifically, our goal is to re-consider what the design of generated interfaces must be to help users adapt their interface to their own needs.

\subsection{Incentivizing Users to Modify Software}

Although users can now modify software in many ways---the content shown, the layout, the style, the available functionality, and the interaction behaviors \cite{chen2025generative, leviathan2025generative, claude2024artifacts}---it is still commonly observed that users do not do it often \cite{mackay1991triggers, alves2026interactiondata}.
A core line of work for interface personalization has explored how to better incentivize users to modify their interface towards one that better suits their needs.

First, users are often unaware of what kinds of modifications they can even make. In traditional end-user software development, users struggle because they lack a robust mental model of how the underlying software system works \cite{ko2011euse, oh2013gesturecustomization, renom2022technicalreasoning, duncker1945problem}, whereas in AI-driven development, users struggle because they also lack an understanding of what the AI model is capable of \cite{zamfirescu2023johnnyprompt, hari2024gulfenvisioning}. A common approach to increasing awareness is by making opportunities to modify the interface more visually salient, either by presenting tooltips \cite{alves2026interactiondata, banovic2012triggering} or by ``anchoring'' customization settings next to the features they modify \cite{ponsard2016anchored}. This helps users modify the interface because it surfaces opportunities for them to \textit{recognize} rather than \textit{recollect} \cite{nielsen1994enhancing}.

Second, users struggle to articulate their own needs and preferences to concrete changes for the interface.
To address this, interfaces can represent customizations to better align with the user's tasks and preferences. For example, users can easily customize data attributes in an interface since these attributes describe the exact criteria they are looking for (e.g., viewing hotels that are dog friendly by surfacing the ``dog friendliness'' attribute) \cite{malleableODI}. Interfaces can also present a design space of potential interfaces the user can create by generating a gallery of variations, enabling the user to find an interface they want without having to specify it from scratch \cite{marks1997designgalleries, lee2010examplegalleries, odonovan2015designscape}.

This suggests that if we can help users (1) become more aware of modification capabilities and (2) articulate their needs and preferences about the interface, we can better incentivize users to modify generated interfaces.
Thus, we aim to generate elicitation techniques in the interface that both reveal opportunities to modify the UI and translate users' intent into concrete changes.

\subsection{Designing Elicitation Techniques to Draw Out User Preferences}

Interactive systems implement elicitation techniques when systems lack the context needed to help the user with their task \cite{horvitz1999principles, gajos2005arnauld, linden1997interactive, pu2003user}.
But every interface is different, tasks are different, and the environments are different, and this all requires designing elicitation techniques that best suit that scenario.
For instance, recommender systems ask users to critique examples to learn what to prioritize \cite{pu2003user, reilly2004dynamiccritiquing}, machine learning tools elicit user input by letting people label examples and correct outputs \cite{olseninteractiveml, dudleyreview}, design space exploration tools offer galleries or canvases of variations to choose from, refining the space toward the user's goals \cite{marks1997designgalleries, odonovan2015designscape, koyama, brochu}, and conversational interfaces pose clarifying questions to specify the user's prompt \cite{tankelevitch2024metacogAI}. 
We have also seen design techniques to elicit user preferences by designing interaction or interface differently, such as by giving users interactive sliders to tune \cite{louie2020novice, priyan2024dynavis}, constraints and filters to select \cite{louie2020novice}, presenting an ambiguous or ``abstracted'' form of the output \cite{gao, cao2024elastica, min2026gradual}, or illustrating user interfaces in wire-frames to afford creative exploration \cite{buxton2007sketchingUX}.




Prior research has also developed elicitation techniques for improving user interfaces for various purposes.
For example, prior systems elicit designer preferences about interfaces to support UI design tasks \cite{swearngin2020scout} as well as to use designer insights for model training \cite{wu2024uiclip, wu2026improving}.
Furthermore, Gajos and Weld have explored how generated interfaces for end-users can elicit preferences both implicitly and explicitly by incorporating example critiquing and comparing two alternatives \cite{gajos2004supple, gajos2005arnauld}.

While these techniques are effective for specific tasks like authoring data visualizations or conversing with AI, GenUI aims to support a broad range of tasks, so a fixed set of techniques cannot guarantee the best fit for any given task.
Instead, our work shares similar perspectives to the argument that we must use AI to generate a broader \textit{design space} of options to structure and diversify what the user can interact with \cite{suh2024luminate, gero2024llmsensemakingscale, malleableODI, meridian, almeda2024dreamsheets}.
Generating and drawing from a design space helps manage the infinite possibilities of what AI can generate while providing the guidance for users and AI systems to select the best option for their needs.
In light of this, we develop Elicitive User Interfaces to map to a design space of elicitation techniques that GenUI systems can draw from to generate the technique best suited to a user's task and preferences.

%% file: sections/04_designspace.tex
\section{Designing Elicitive User Interfaces} 
\label{4_design_EUI}

Elicitation can take many forms depending on what the system needs to elicit and how it asks the user. To understand this, we first examine where preferences about generated interfaces go unexpressed, then develop a design space for how GenUI systems can elicit them.

\subsection{Understanding What Users Want to Change but Do Not Ask For}

To understand how users currently express preferences in GenUI, we conducted a formative study in which participants reacted to generated interfaces. We organized the study around two questions. 

\begin{itemize}
    \item \textbf{[RQ1]} How do users currently experience modification opportunities in GenUI? 
    \item \textbf{[RQ2]} When users do want to modify, what kinds of changes do they want? 
\end{itemize}

\subsubsection{Procedure}
We recruited ten participants through an online community at our institution. Participants evaluated interfaces generated in advance using a commercial GenUI tool (Claude Artifacts~\cite{claude2024artifacts}), allowing us to show the same stimuli to everyone. Each scenario contained three to four chat-to-UI turns across three tasks including learning about credit scores, drafting a cover letter, and planning an event under budget and food constraints. The interfaces used components such as gauges, checklists, timelines, ranked lists, and cost breakdowns. The full stimuli are reported in Appendix~\ref{app:formative-stimuli}.

For each interface, participants identified what they would keep or change, what they would prefer instead, and whether they would actually make the change. They could point to, sketch, or annotate specific interface regions. We recorded the requested change, its rationale, and whether participants could specify an alternative. The study was approved by our institution's IRB, and each participant was compensated approximately USD 20.

\subsubsection{Formative Study Results}
\begin{figure*}[h]
    \centering
    \includegraphics[width=\linewidth]{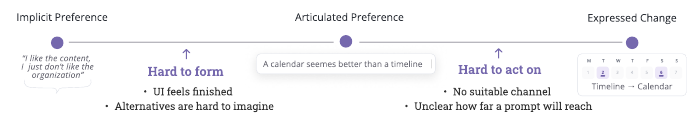}
    \caption{Where UI preferences can go unexpressed. Participants had difficulty both forming preferences and acting on preferences they could articulate.}
    \Description{A three-stage diagram showing how a preference moves from implicit
to articulated to expressed. The first stage, ``Implicit Preference,'' includes
the example, ``I like the content, I just don't like the organization.'' An
arrow labeled ``Hard to form'' leads to ``Articulated Preference,'' illustrated
by the statement, ``A calendar seems better than a timeline.'' Reasons that
preferences may be hard to form include that the UI feels finished and that
alternatives are hard to imagine. A second arrow, labeled ``Hard to act on,''
leads to ``Expressed Change.'' Barriers to acting include having no suitable
channel and being unsure how far a prompt will reach. The final stage shows the
interface changing from a timeline to a calendar.}
    \label{fig:placeholder}
\end{figure*}

\paragraph{When prompted, users identified many situated preferences about the interface.}
Participants produced 332 requests for changes when prompted to consider them. Within the same UI component, participants requested different modifications. For a budget pie chart in the planning task, P1 asked for drill-down interaction (\say{a pie chart inside the pie chart}), P8 wanted it removed in favor of a plain list because \say{what to buy matters, not percentages,} and P9 rejected the component altogether, since the ratio did not matter to him. Although the specific changes varied, they fell into a set of recurring types. Participants wanted to change a component's \emph{behavior} (123/332), \emph{representation} (58), \emph{density} (52), \emph{form} (42), or \emph{emphasis} (23), with another 34 requests concerning whether a component existed at all. Participants wanted to add, remove, replace, or adjust these aspects. For example, P5 wanted to add interaction by asking to \say{input my own values and see the result}, while P3 wanted to replace a timeline with a calendar. Thus, prompting participants to reflect can surface preferences about the current interface, even when they had not previously expressed them.

\paragraph{Users could not articulate their needs when the interface felt finished or alternatives were hard to imagine.}
Participants often mentioned a polished and completed appearance as a reason why they did not consider changing the interface and articulating their needs. P2 explained that shadows and rounded edges were \say{the format I'm used to seeing in a finished product}, so his first instinct was not \say{I can change this}. In contrast, \say{if it had given me more of a pre-UI, just plain boxes, flat design, then I'd think this is something I can change right now}. P3 similarly worried that changing an already attractive interface through an imprecise request might make it worse. Finished interfaces also made alternatives harder to imagine. P4 described how \say{having already seen it, my thinking converged on it---nothing better comes to mind}, noting that richer visualizations constrained his imagination more than plain text.  

\paragraph{Users could articulate some preferences but did not express them when the scope of a change was unclear.}
Even when participants could state what they wanted, they did not always attempt the change. Participants were concerned that a prompt could not be contained to the part of the interface that they intended to change. P2 said he was \say{afraid that it would not know where `here' is and change the design of everything}, concluding that he would \say{just target big things}. P4 similarly wanted direct editing for small changes and scoped AI requests for larger ones. P1 made the scoping problem explicit. Image editors let him change \say{exactly the part I select}, but after asking the system to append to an interface and receiving a rebuilt one, he concluded that interface editing was simply unavailable.

\subsection{A Design Space of Elicitation Techniques for GenUI}

\subsubsection{The Need for Elicitation in GenUI}
Our formative study showed several ways users might not express preferences about a generated interface. When prompted to reflect, participants could identify many changes they preferred. Yet, some preferences were difficult to form, while others could be articulated but were not acted on. Participants themselves suggested ways the interface could help such as showing a rough version (P2), presenting alternatives (P6), asking  before generating (P8), or letting users indicate which part of the interface they wanted to change (P1). These observations suggest that the interface itself can help users draw out preferences that might otherwise remain unexpressed.

However, there is no single design for elicitation that can fit every situation. Whether an elicitation was useful depends on what was being decided, when it appeared, and the user encountering it. We therefore consider a space of ways that GenUI can ask the user. This motivates our design space of \textbf{\textit{Elicitive User Interfaces (EUIs)}}---interface elements and generation strategies that expose decisions and invite users to shape them as the interface is generated. What are the ways such generated elicitations can vary so that a GenUI system can be instructed to design the right one in the right place?

\subsubsection{Derivation Method}
We developed the design space from three sources. First, our formative study identified different reasons preferences remain unexpressed, as well as elicitation approaches suggested by participants. Second, we drew on elicitation and uncertainty-signaling techniques outside GenUI, such as clarifying questions, preference elicitation in recommender systems and design galleries~\cite{marks1997designgalleries}, and conventions for communicating uncertainty through interface design. Third, we examined previous GenUI and HCI systems that let users modify generated output~\cite{priyan2024dynavis, jelly}.

Because there is no established design practice for elicitation in GenUI, we began by exploring the space through design. Two authors generated candidate elicitation techniques that addressed the barriers observed in the formative study, implementing working probes. We then compared these techniques to identify the underlying design choices that distinguished them. When techniques captured the same design choice, we merged them and when a technique could not be described by the existing dimensions, we revised the space. Through this iterative process, we arrived at six axes that distinguish the techniques in our design exploration. 



\begin{figure*}[h]
    \centering
    \includegraphics[width=1\linewidth]{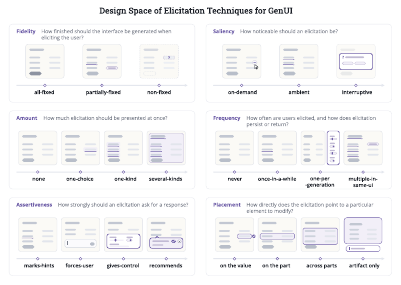}
    \caption{Design space of elicitation techniques for GenUI, organized along six axes---fidelity, saliency, amount, frequency, assertiveness, and placement.}
    \Description{A design space of elicitation techniques for generative user interfaces organized along six axes--- fidelity, saliency, amount, frequency, assertiveness, and placement. The axes describe how finished the interface appears during elicitation, how noticeable the elicitation is, how much elicitation is presented at once, how often it appears or returns, how strongly it asks for a response, and how locally it is attached to the interface.}
    \Description{A design space of elicitation techniques for generative user
interfaces, organized into six axes. Each axis is shown with example interface
illustrations arranged along a continuum. ``Fidelity'' describes how finished
the interface is when eliciting the user, ranging from all-fixed to
partially-fixed to non-fixed. ``Amount'' describes how much elicitation is shown
at once, from none to one-choice, one-kind, and several-kinds. ``Assertiveness''
describes how strongly the interface asks for a response, ranging from
marks-hints to forces-user, gives-control, and recommends. ``Saliency''
describes how noticeable the elicitation is, from on-demand to ambient to
interruptive. ``Frequency'' describes how often elicitation appears or returns,
from never to once-in-a-while, one-per-generation, and multiple-in-same-UI.
``Placement'' describes how directly the elicitation points to what can be
modified, ranging from on the value to on the part, across parts, and artifact
only.}
    \label{fig:placeholder}
\end{figure*}

\subsubsection{The Six Design Axes of Elicitive User Interfaces}


\paragraph{Fidelity}
\textit{How finished should the interface be generated when eliciting the user?}
At one end, parts of the interface that are open to change can be shown as unfinished, and at the other, elicitation can occur within a fully rendered interface. Our formative study suggested that by being visually polished, the interface felt unmodifiable to users and discouraged them to consider alternatives. Lower fidelity can instead signal that a design remains open to change, just as rough prototypes have been used to invite feedback~\cite{dow2011parallelprototyping, buxton2007sketchingUX}. In GenUI, this could take the form of a wireframe before generation or unfinished elements within an otherwise finished interface.

\paragraph{Saliency}
\textit{How noticeable should an elicitation be?}
Saliency ranges from visually quiet marks that users may discover while inspecting the interface to prominent elicitations that draw attention. Whereas fidelity signals whether the interface remains open to change, saliency indicates how strongly an opportunity for input is made visible. We can see this range in conventional interfaces---a comment indicator can remain subtle until hovered, a suggested edit can be highlighted inline, while an onboarding overlay can demand attention before the user continues. If an elicitation is greater in salience, it makes it harder for the user to miss it but it also makes the user put more attention to it.

\paragraph{Amount}
\textit{How much elicitation should be presented at once?}
Amount describes how much the interface asks about at once. Our formative study suggested that seeing possibilities could help users form preferences. Amount can therefore vary through the number of questions, alternatives, or undecided regions presented together, as in design galleries that expose multiple possibilities for comparison~\cite{marks1997designgalleries}. More elicitation gives the user more opportunities to react to, but also increases how much they must consider at once.

\paragraph{Frequency}
\textit{How often are users elicited, and how does elicitation persist or return?}
Frequency describes whether and when elicitation repeats over repeated interaction with an interface. An elicitation might appear only once, remain until it is answered or dismissed, or return when changes to the artifact make the decision relevant again. Participants also differed in how long they wanted to keep refining an interface. For artifacts they would use only briefly, they saw little reason for the system to keep asking about changes. GenUI can therefore vary not only how often it asks, but what happens to an elicitation over time---whether it disappears, remains available, or returns after the interface changes.

\paragraph{Assertiveness}
\textit{How strongly should an elicitation ask for a response?}
Assertiveness describes how strongly the system pushes the user to resolve the decision it exposes. At a low level, an elicitation may simply make a decision available to see for the user, but more assertive forms can proactively raise a question or pause generation until the user responds. In the formative study, P8 wanted the system to ask about her situation before generating rather than waiting for her to identify what information was missing. The same underlying decision could therefore be surfaced as an optional hint, an explicit question, or a gating question that must be resolved before generation continues~\cite{tankelevitch2024metacogAI}.

\paragraph{Placement}
\textit{How directly does the elicitation point to a particular element to modify?}
Placement describes the scope to which an elicitation is attached, from a particular value or element, to a component or region, to the artifact as a whole. Our formative study suggested that this grounding could make the expected reach of a change clearer. P2 was \say{afraid that it would not know where `here' is and change the design of everything}, while P1 wanted to change \say{exactly the part I select}. Elicitation can therefore be placed directly on a value or component, attached to a larger region, or presented globally. Existing steering and direct-manipulation tools similarly use this locality to scope interaction~\cite{priyan2024dynavis}.

\subsubsection{Limitations}
We intend this space as a starting vocabulary rather than a complete taxonomy. It was derived from a formative study and our own design exploration, and we hope that future systems and research can refine and extend its axes as new forms of elicitation emerge.

%% file: sections/05_probe.tex
\section{Generating Elicitive User Interfaces} 
\label{5_gen_EUI}


Our design space describes what an elicitation can be. To generate EUIs, however, a GenUI system must also decide what is worth eliciting and how that decision should be elicited. This is particularly important for GenUI because it contains many decisions that were not specified in the user's request. We describe how we use the design space to make these decisions during generation. 

\subsection{Integrating the Design Space into EUI Generation}
In our first attempts, we treated the design space as a catalogue of techniques and asked the model to select among them. The model often selected techniques without considering what the underlying decision was about, sometimes producing mismatches, such as attaching a question about the user's situation to a layout choice. We also tried mapping specific scenarios to techniques, but these mappings did not generalize well to new requests. More general rules written in prose had little effect on what the model selected. 

What worked better was treating the axes as parameters the model has to decide on when generating an EUI. Given the user's request and the initially generated interface, the pipeline first identifies what is worth eliciting and makes a decision along each relevant axis. For example, it considers how broadly a change would affect the interface (placement), how strongly to ask for a response (assertiveness), and whether to ask before the interface is fully rendered (fidelity). These decisions determine where the elicitation falls within the design space and guide which technique is used. 



\subsection{Instantiating the Design Space in \sysname{}}
We instantiate this generation mechanism in \sysname{}, a design probe that generates working interfaces with EUIs embedded within them. \sysname{} also includes a conventional prompt box, allowing users to request changes directly when they are not surfaced by an EUI. The \sysname{} interface is reported in Appendix~\ref{app:Ditto}.

\subsubsection{Elicitation Techniques Supported in \sysname{}}
\begin{figure*}[h]
    \centering
    \includegraphics[width=\linewidth]{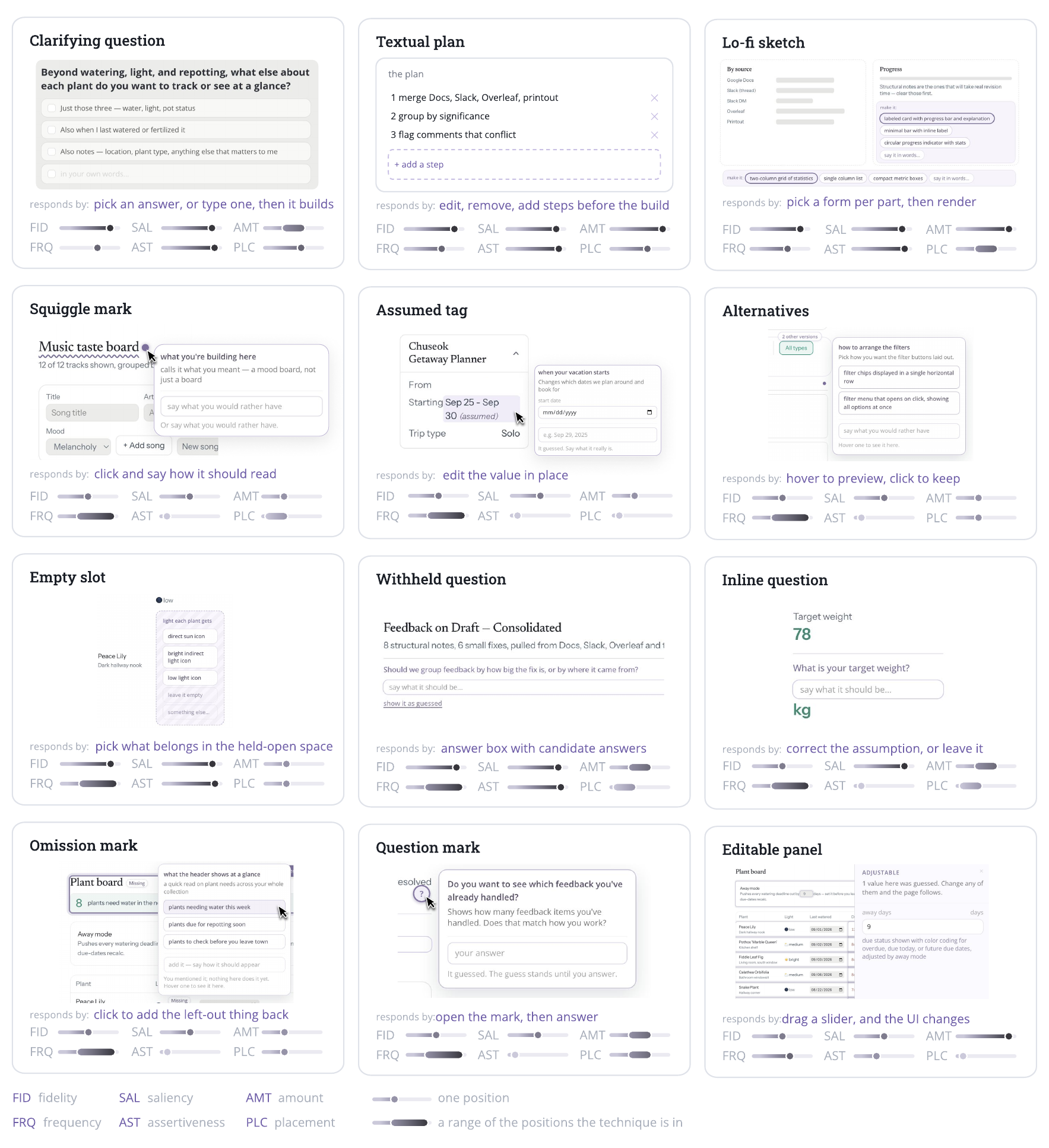}
    \caption{The twelve elicitation techniques implemented in \sysname{}, spanning different positions in the EUI design space.}
    \Description{Twelve elicitation techniques implemented in \sysname{}, spanning different positions in the EUI design space. Each technique is shown with its corresponding positions along the axes in the design space and the interaction through which users can respond.}
    \label{fig:ditto-techniques}
\end{figure*}

Of the designs explored while constructing the design space, \sysname{} supports twelve techniques spanning different positions along its axes. These range from techniques that appear before generation, such as clarifying questions, textual plans, and
lo-fi sketches, to techniques embedded in the generated interface, such as question marks, empty slots, alternatives, and editable panels. Each technique represents a combination of positions across multiple axes. Each technique also pairs a way of surfacing a decision with a fixed response control. For example, an \textit{assumed tag} marks a value the system inferred but the user would know, such as a weekly budget, with a subtle tint in the finished interface. The user can respond by editing the value directly in place. A \textit{withheld question}, by contrast, asks the user to provide input before continuing, using an answer box and candidate answers as its response control. Figure~\ref{fig:ditto-techniques} shows the twelve techniques, their positions in the design space, and their response controls.

\subsubsection{Elicitation Pipeline}


\sysname{} generates EUIs in four steps.

\paragraph{1. Set task-level constraints.}
Before generation, \sysname{} considers four properties of the task--- whether the user is likely to revisit the artifact, how consequential an incorrect assumption would be, whether different users would likely want different artifacts, and whether the user could have specified what they wanted without seeing it. These properties constrain when and how elicitation should occur, rather than selecting a particular technique.

\paragraph{2. Find decisions and position their elicitations.}
\sysname{} first generates the interface without elicitations. It then identifies decisions made during generation, such as ordering, assumed user preferences, and omitted information. These decision types were informed by the kinds of changes participants
made in our formative study. Decisions already specified by the user or supported by a clear convention are removed. For the remaining decisions, \sysname{} determines where an elicitation should fall along the relevant axes in the design space.

\paragraph{3. Validate and render.}
The resulting axis positions are checked against the task-level constraints before elicitations are attached to the relevant interface elements. \sysname{} generates their wording, while each technique's visual appearance and response controls are fixed. 

\paragraph{4. Retire answered elicitations.}
Once a user responds through an elicitation, the corresponding decision is no longer treated as unresolved and the elicitation disappears. New elicitations can appear when subsequent requests introduce new decisions.

\subsubsection{Implementation Details}
\sysname{} consists of a Python server and React client, with generated interfaces running in a sandboxed frame. We use \texttt{Claude Sonnet 5} for interface generation and modification and \texttt{Claude Haiku 4.5} for identifying candidate decisions and positioning them along the design space axes. We provide the design space definitions, technique specifications, and generation instructions as a skill package, along with the \sysname{} probe, in the supplementary material.

%% file: sections/06_userstudy.tex
\section{User Study 1: Experiencing Elicitive UIs}

To understand how users experience EUIs, we conducted a user study with the following questions. 

\begin{itemize}
    \item \textbf{[RQ1]} How do EUIs surface preferences that might otherwise remain unformed or unexpressed?
    \item \textbf{[RQ2]} How do the design axes of EUIs shape users' responses to elicitation? 
    \item \textbf{[RQ3]} How are responses to elicitation shaped by the user and task?
\end{itemize}

\subsection{Procedure}

\subsubsection{Participants}
We recruited 12 participants through an online community at our institution. Eight had formal education in design or frontend development, ranging from coursework to professional practice, four had neither, from complete novices to a self-taught developer. More detailed information on the participants is reported in Appendix~\ref{appendix:participants}. 

\subsubsection{Procedure}
Each study session lasted approximately 120 minutes. It began with a 15-minute introduction and demonstration of \sysname{}, during which participants were introduced to all supported elicitation techniques. Then, participants went through three 25-minute sessions. For the first two sessions, participants used the standard \sysname{} pipeline, in which the system selected which elicitation techniques to present. Before the third session, participants were given the option to manually select the techniques. After each session, participants took 10 minutes to rate the techniques they were exposed to. After all three sessions, they completed a survey asking about the overall experience. The study was conducted with approval from our institution's IRB, and each participant was compensated approximately USD 20.

\subsubsection{Tasks}
Each participant completed three tasks in learning, writing, and planning. We chose these task types based on our formative study, which suggested that opportunities for elicitation could differ with properties such as how subjective the desired result is, how long the artifact will be used, and how easily users can specify what they want upfront.

Participants completed the three tasks in counterbalanced orders, with three participants assigned to each ordering. Within each task type, participants chose the specific task they wanted to complete. For example, for the learning task, participants were asked to choose a concept or topic they wanted to learn about and prompt \sysname{} accordingly (e.g., \say{Teach me how to prepare food from different countries.}). We report the full task instructions and participants' specific prompts in Appendix~\ref{tab:queries}.

\subsection{Measurements}
\paragraph{Quantitative Measurements}
For each session, we logged the participant's initial query and follow-up prompts, the generated interface and its components, and every elicitation mark shown. A mark is one component's elicitation in one session. For each mark, we recorded its technique, position along the design-space axes, and whether the participant acted on it by selecting an option, answering a question, or entering a response. 

\paragraph{Qualitative Measurements}
Participants thought aloud throughout each session. After each session, they reviewed the elicitation marks they had encountered and rated how glad or annoyed they were that each mark appeared, how useful they found it, and whether the mark surfaced a preference that was already on their mind or emerged from seeing the mark. For each technique, participants also indicated whether they would want more, the same, or fewer such marks in a similar task. At the end of each session, they rated how well the final interface matched what they needed, and after the final session, they completed a short closing survey. 

\subsection{User Study 1 Results}

We report the findings by each research question. In summary, EUIs helped users surface preferences that they would have otherwise not formed or expressed, while the axes in the design space shaped different aspects of the elicitation experience, and users had specific preferences for each of the axis placements. 


\subsubsection{RQ1. How do EUIs surface preferences that might otherwise remain unformed or unexpressed?}

\begin{figure}
    \centering
\includegraphics[width=\linewidth]{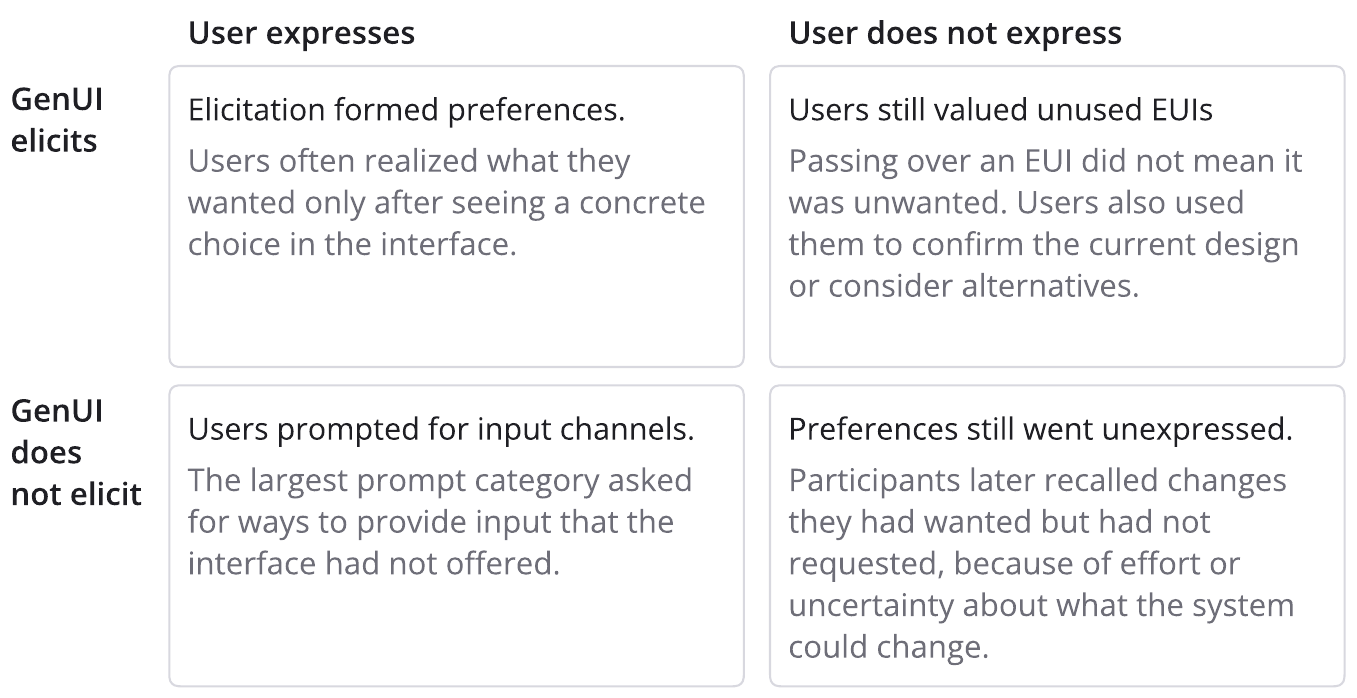}
    \caption{RQ1---How preferences surfaced depending on whether the GenUI elicited them and whether users expressed them.}
    \Description{A two-by-two matrix organizing how preferences surfaced in Study 1. The columns distinguish whether the user expressed a preference, and the rows distinguish whether the GenUI elicited it. When the GenUI elicited and the user expressed, elicitation often helped form preferences. When the GenUI elicited but the user did not express, unused EUIs could still be welcome by helping users confirm the current design or consider alternatives. When the GenUI did not elicit but the user expressed, participants often asked for additional input channels. When neither occurred, some preferences remained unexpressed because of effort or uncertainty about what the system could change.}
    \label{fig:rq1-matrix}
\end{figure}

We organize our observations along two dimensions: whether the system elicited the preference, and whether the participant expressed it (Fig~\ref{fig:rq1-matrix}).

\paragraph{Most expressed preferences emerged in response to elicitation.}
Across the 36 sessions, the system showed 330 elicitation marks, and participants eventually acted on 139 of them (42\%), by answering embedded questions, choosing among alternatives, or filling held-open slots. Of the rated preferences ($n=169$), 63\% were reported as emerging from \say{seeing the mark}, rather than being \say{already on your mind}. Thus, many of the preferences were not already formed before the interaction, but emerged as the EUI presented something to react to. P1 explained, \say{When you see the UI first, it's somewhat already usable, standardized. So you can't really think of what you would need more. The marks suggest what you would have missed.} Similarly, P2 had not considered a comparison component beforehand, but \say{seeing it, [I] realized that you can.} 

\paragraph{EUIs enabled users to act where a prompt box did not.}

Participants' accounts pointed to three reasons they acted through EUIs when they otherwise might not have prompted. First, EUIs surfaced possibilities participants would not necessarily have considered themselves (P2, P3, P4, P6, P12). P12 made this particularly explicit---\say{If I hadn't seen the alternatives and possible changes, I would probably have been satisfied and just used it---being exposed to these possibilities makes me interact.} 

Second, participants described the marks as making changes feel safer or more predictable (P2, P5, P7). P2, who said she normally did not prompt for UI changes even in Claude Artifacts, explained that \say{in this case I could, because it shows you the types of decisions that go on behind the scenes when making UIs.} She described this as forming a belief that the system could \say{pull this off}, and later said the techniques increased both the specificity and frequency of her prompts. 

Finally, participants repeatedly treated marks as a way to scope a change to a particular part of the interface (P1, P3, P4, P6). P6 summarized the distinction as: \say{Wanting to change the whole idea---that's a prompt thing. A mark is changing a local thing.} 


\paragraph{Without elicitation, some preferences were prompted and others went unexpressed.}

When participants wanted to express something beyond what the marks asked, they could still use the prompt box. Across the sessions, participants made 54 unique prompts. The largest category (17 of 54) requested input affordances, such as surveys, comment fields, filters, toggles, or editable text, rather than directly specifying a change to the UI. P8, for example, asked, \say{Could you start with a general survey that asks critical questions about my crop of choice, the country I live in\ldots} In these cases, participants used prompting not to specify a preference directly, but to ask the system for another way to provide input. Some participants even prompted for the elicitation mark itself. P3 ended one session by asking, \say{please add mark tools especially}, while P12 \say{wanted to prompt for more marks but didn't know how to prompt for them}.

Prompting did not capture everything participants wanted when the interface did not elicit. We identified eleven preferences that participants wanted but never expressed through any interaction. P4, for example, thought \say{tabs 2, 3, 5 could be merged\ldots} but was \say{too lazy to edit that}, while P3 recalled, \say{[I] wanted different color differences in the medium' and low'---but I didn't ask for it.} These barriers align with the formative study participants did not know a change was possible, decided it was not worth the effort, or doubted that the system could make the change. Thus, when an EUI did not surface a decision, prompting only partially filled the gap. 

\paragraph{Participants still found unused EUIs useful.}

Not acting on a mark did not mean participants found it unhelpful. Marks that participants rated but did not act on were rated almost identically in how welcome they felt compared with marks they acted on (3.97 vs.\ 3.98 of 5), and only slightly lower in usefulness (3.92 vs.\ 4.02; $n=125$ and $119$ rated marks, respectively). Participants described two reasons for this. Sometimes, the existing UI was already good enough (P2, P4, P7, P12). P2 described one mark as \say{not useful, because the original UI itself was good enough}, while still being \say{glad it was there\ldots more space for my freedom}. Seeing an alternative gave participants an opportunity to confirm that they preferred what was already there. In other cases, the mark was useful because it prompted participants to consider possibilities without necessarily choosing one given by the EUI (P3, P6, P10). As P6 explained, \say{although you don't actually use the options, it still triggers you to think---what else could there be?} These results show that interaction alone is an incomplete signal of an elicitation's value.


\subsubsection{RQ2. How do the design axes of EUIs shape users' responses to elicitation?}
The six axes affected different aspects of how participants responded to elicitation (Table~\ref{tab:rq2-dimensions}).

\begin{table}[t]
  \centering
  \caption{RQ2---How the six EUI design axes shaped participants' experiences with elicitation.}
  \Description{A table summarizing how the six EUI design axes shaped participants' experiences with elicitation. For each axis---saliency, fidelity, amount, frequency, assertiveness, and placement---the table reports the main observed effect and a qualification describing when that effect varied by participant, task, or context.}
  \label{tab:rq2-dimensions}
  \small
  \begin{tabular}{p{0.16\linewidth} p{0.34\linewidth} p{0.40\linewidth}}
    \toprule
    \textbf{Axis} & \textbf{What we observed} & \textbf{Qualification} \\
    \midrule

    Saliency
      & More salient marks were acted on more often (50\% vs.\ 37\%)
      & Both were similarly welcome; preferences for more salient marks differed by participant \\

    Fidelity
      & Lower fidelity made the UI feel more open to change 
      & Preferred fidelity depended on the task and stage \\

    Amount
      & More elicitation did not consistently increase engagement
      & Participants cared more about what was asked than how much was asked \\

    Frequency
      & Marks were acted on more often when first shown than when repeated (24\% vs.\ 13-15\%)
      & Participants wanted marks to return in some situations, rather than simply more or less often \\

    Assertiveness
      & Participants generally preferred marks that proposed alternatives over marks that only signaled uncertainty
      & How proactive they wanted the system to be differed by participant \\

    Placement
      & Placement helped communicate what part of the UI a mark referred to
      & Appropriate placement depended on the decision being elicited \\

    \bottomrule
  \end{tabular}
\end{table}


\paragraph{Saliency affected whether participants acted on a mark, and fidelity affected how open the interface felt to change.} Participants acted on 50\% of interruptive marks (67 of 134), compared to 37\% of ambient marks (72 of 195). P1 described an interruptive mark as something that \say{feels like a part of the UI instead of something decorative}, adding that this could make it more annoying but also make them \say{answer it fast}. P4 similarly said that more salient marks could be \say{more annoying}, but that they would \say{probably answer more}, noting that they had missed many of the less visibly salient question marks. Fidelity affected how open the interface felt to change. Low-fidelity sketches had the second-highest action rate among the techniques we tested (47\%, 38 of 81 shown). Participants described roughness as signaling that the interface was still open to revision. P12 explained that \say{lofi gives the feeling that it's easier to change and fix---an unfinished state is psychologically easier to overturn}, while P1 felt that \say{in fully finished UIs there is no freedom}. 

\paragraph{Frequency shaped how participants expected elicitation to behave over time.} Of the 330 marks, participants acted on 80 at their first showing (24\%), and among marks shown again after being passed over, action rates fell to 13--15\%. Participants' accounts, however, suggest that this did not simply mean they wanted fewer marks. They distinguished between repeating an existing ask and introducing elicitation in response to something new. P7 described some marks as \say{placeholder marks that later on I would switch off for the final UI}, while P1 said that \say{if they had asked with the same strength at the end as the initial stages, it would have been annoying}. In contrast, when a mark appeared in response to an edit P7 had just made, he said \say{it makes you realize the UI is like a living state}, with three participants also asking for more marks late in their sessions. P5 described both expectations for the same mark: \say{I want them to disappear after I prompt---but if I don't like the output, having them still there could be better\ldots this is conflicting.} Thus, frequency concerned not only how often users were asked, but when an elicitation should persist, disappear, or return as the interface changed. 

\paragraph{Other axes shaped the interaction in different ways.} EUIs that were more propositional were generally preferred to EUIs that only pointed out uncertainty. In the post-session technique votes, Alternatives received the most requests to see more (9 more, 0 drop), whereas Squiggle received the most requests to remove it (6 drop among 11 votes). This preference was not uniform in our study---while several participants found stronger forms of elicitation intrusive, P12 wanted the system to \say{have asked me more} and to have been \say{more proactive}.

Amount showed less consistent effects on users. Participants often evaluated amounts in terms of whether the individual questions were worth asking. P4 felt that some questions were \say{asking too obvious things}, while P12 summarized their preference as \say{need better questions, not more questions}. Placement helped participants understand the scope of an ask. Participants described EUIs attached to particular components as making clear what part of the interface was being changed (P1, P3, P4, P6). As P6 explained, \say{it's on this location, and it gives context to the agent about what part I want to change}.

\paragraph{The task decided when to elicit, and how settled the UI should look.}
Task-related differences were most apparent in when elicitation appeared and the fidelity of the interface around it. Participants often described earlier elicitation as useful when the task still left consequential decisions open. P1, for example, preferred elicitation earlier in a planning task because \say{aligning big parts in the initial stages, where there is a lot of freedom, is easier\ldots I wouldn't want to see a full-fidelity UI for this}. Conversely, P10 questioned an ask that appeared after the relevant decision had already been made in a writing task. \say{it feels like an important question, but why ask now? It could have been asked in the previous stages}. Fidelity was similarly experienced in relation to what the task required at a particular stage. P5 regretted using low fidelity for a content-heavy task, explaining that \say{this one is really about the information\ldots actually regret putting in lofi}, whereas P4 preferred an earlier, less settled interface for planning. Thus, task affected more of when elicitation should happen, rather than which kind. 



\subsubsection{RQ3. How are responses to elicitation shaped by the user and task?}


\paragraph{Responses towards EUIs varied more by person than by task.}

\begin{table}[t]
  \centering
  \caption{Share of variance in each measure explained by participant and task ($\eta^2$), computed over 36 sessions from 12 participants. Participant identity explained more variance than task across all four measures.}
  \Description{A table comparing the share of variance in four study measures
explained by participant identity and task. Across all measures, participant
identity explains more variance than task. The participant and task
$\eta^2$ values are .43 and .04 for welcome of being asked, .48 and .03 for
engagement rate, .47 and .05 for page satisfaction, and .35 and .15 for typed
versus clicked input, respectively.}
  \label{tab:rq3-person-task}
  \begin{tabular}{lcc}
    \toprule
    Measure & $\eta^2$ person & $\eta^2$ task \\
    \midrule
    Welcome of being asked      & .43 & .04 \\
    Engagement rate             & .48 & .03 \\
    Page satisfaction           & .47 & .05 \\
    Typed vs.\ clicked input    & .35 & .15 \\
    \bottomrule
  \end{tabular}
\end{table}

To examine whether responses were shaped more by the person or the task, we compared how much of the variation in each measure could be attributed to participant versus task. Across the 36 sessions (12 participants $\times$ 3 sessions), we quantified this using $\eta^2$, the share of variance explained by each grouping.

From our analysis, participant identity accounted for more variance than task across all four measures (Table~\ref{tab:rq3-person-task}): how welcome being asked felt ($\eta^2=.43$ vs. $.04$), engagement ($.48$ vs. $.03$), satisfaction with the UI ($.47$ vs. $.05$), and whether participants responded by typing or clicking ($.35$ vs. $.15$). Although none of these tests reach conventional significance ($p$ = .057–.64), they show a consistent direction that aligns with what we observed across the sessions. P3 liked the EUIs but rarely acted on them, saying \say{even though I don't interact with it, I like the idea of having it.} In contrast, P6 acted on the EUIs through prompting rather than the provided clicks or alternatives, but still rated them valuable because \say{they were useful suggestions that you can choose to ignore}.


\paragraph{Elicitation preferences showed more consistency than UI preferences.}
We compared both the UI choices participants made and their preferences toward the elicitation techniques. For the latter, we constructed an \textit{elicitation profile} for each participant in each session, capturing how much they liked each technique and which techniques they wanted to see more of or drop. We then compared profiles using rank correlations to see whether a participant's profile was more similar to their own across sessions than to those of other participants.

At the UI level, we found little evidence of preferences that could be carried across sessions. We attempted a simple analysis by aligning every modification choice a participant made across their sessions and asking whether the same person chose the same way twice. When comparable choices appeared, participants did not reuse the same UI forms more often than in the shuffled comparison (self-reuse $.075$ vs. $.139$). For instance, P1 chose a weekly calendar for one page and a badge list for the next. Direct comparisons were also sparse because each session produced a different page for a different task, so a participant almost never faced the same decision about the same kind of component twice. There were only six comparable pairs across the entire study and 85\% of choices had no counterpart in their other sessions at all.

Participants' choices also appeared closely tied to what the current interface presented. At elicitation marks, 79\% of preference actions selected one of the alternatives shown by the system, and typed responses often referred directly to the current page. Participants could articulate broader preferences when asked, such as preferring \say{clean layouts} (P10), but these did not specify what the interface should do in a particular context. Within three sessions, we therefore could not identify UI preferences that were stable and specific enough to reuse across contexts.

In contrast, we observed greater consistency in how participants preferred to be elicited. This appeared both in their overall receptivity to being asked and in their preferences for particular EUIs. A participant's average receptivity in the first two sessions predicted their receptivity in the third ($\rho=.69$, $p=.045$). Participants' ratings of which techniques felt welcome were more similar to their own earlier ratings ($+.20$) than to those of other participants ($-.01$; $p=.062$), as were their more-or-drop votes ($+.49$ vs. $+.19$; $p=.082$). Participants also began to articulate more specific rules about how they wanted to be asked. P2 preferred embedded question cards over quiet question marks for text-heavy tasks \say{because it makes you think about deeper things, and a question mark would not be visible among so much text}. Seven of twelve participants articulated rules of this kind by their second or third session. This suggests that compared with their preferences for particular UI forms, participants developed more consistent preferences for how they wanted to be elicited.

%% file: sections/07_longstudy.tex
\section{User Study 2: Elicitation Across Sessions}  

Study 1 suggested that responses to elicitation varied more across users than across tasks, and that participants began to develop preferences for how they wanted to be elicited as they gained experience with EUIs. However, each participant encountered \sysname{} only within a single study session. We therefore conducted a three-day deployment with returning participants. We set the following questions.

\begin{itemize}
    \item \textbf{[RQ1]} How do users' elicitation preferences develop across repeated use?  
    \item \textbf{[RQ2]} How do users interpret and want to control a model of their elicitation preferences?
\end{itemize}

\subsection{Participants and Procedure}

We recruited three participants from Study 1 (P2, P3, P8) for a three-day deployment of \sysname{}. Participants used the system for their own tasks in at least six sessions, each lasting at least 25 minutes. After each session, they completed a survey reporting their experiences with the EUIs they encountered. At the end of the deployment, participants took part in a 20-minute remote interview. The study was conducted under approval from our institution's IRB, and the three participants received an additional USD 70.  We report additional study materials and measures in Appendix~\ref{tab:study2-queries}.  

For the first three sessions, participants used the original \sysname{} pipeline without a user-specific elicitation profile. We then constructed a profile for each participant from their EUI interactions and survey responses. Beginning with the fourth session, the pipeline used this profile to guide EUI generation, and the profile was updated after each subsequent session. In the final interview, we showed participants their resulting profile, estimated from their behavior and survey responses.

\subsection{User Study 2 Results}
P2, P8, and P3 completed six, six, and seven sessions over three to four days, respectively. Participants chose their own tasks, which varied substantially across and within participants. These included an interface comparing how NLP and HCI papers argue and a study interface for the four Gospels (P2), a pitch-writing worksheet for a fifth grader and a worst-case projection of the US bond market (P8), and a pastry curriculum and a long letter to a friend (P3).

\begin{figure}
    \centering
    \includegraphics[width=0.85\linewidth]{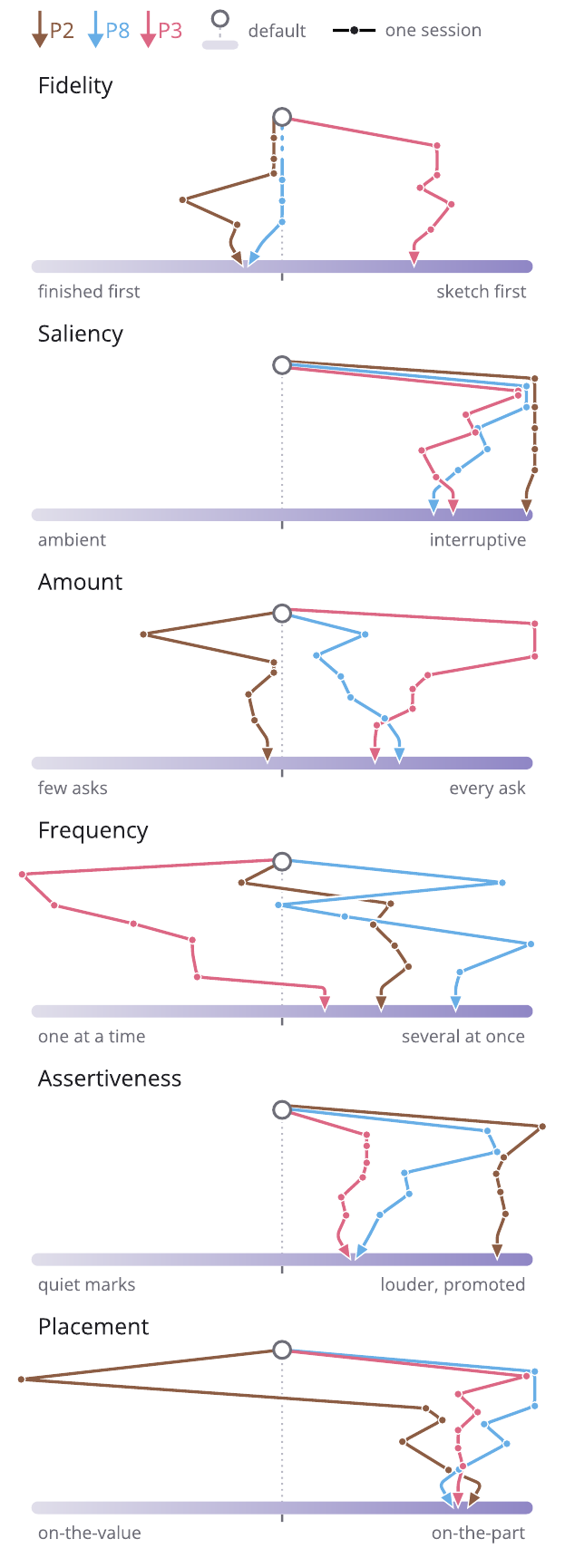}
    \caption{Elicitation profiles for P2, P8, and P3 across repeated sessions. Each panel shows one axis, with session-by-session profile estimates.}
    \Description{Six panels show how the elicitation profiles of P2, P8, and P3 changed across sessions for Amount, Assertiveness, Proactivity, Saliency, Fidelity, and Placement. In each panel, the three participants begin at the midpoint and move downward across sessions toward their final profile value; dots mark observed sessions, dotted segments indicate periods without data, and bars indicate the setting ultimately applied by the system. Amount ranges from few asks to every ask; Assertiveness from quiet to more prominent elicitation; Proactivity from asking first to acting with undo; Saliency from ambient to interruptive; Fidelity from finished first to sketch first; and Placement from on-the-value to on-the-part. The participants differed most on Amount, Assertiveness, Proactivity, and Fidelity, while their profiles were more similar on Saliency and Placement.}
    \label{fig:study2-results}
\end{figure}


\subsubsection{RQ1. How do users' elicitation preferences develop across repeated use?}

\paragraph{Several elicitation preferences stabilized quickly, while others continued to vary with context.}
We estimated each participant's profile using their first $k$ sessions and recorded when each coordinate stopped changing (Figure~\ref{fig:study2-results}). For all three participants, several profile coordinates settled by the second session and did not change afterward. Fidelity, however, changed later for both P8 (session 7) and P3 (session 5). Their responses help explain this difference. Whether they wanted to respond to a rough sketch depended on what the sketch represented. This follows the pattern we saw in Study~1, where participants described preferred fidelity in relation to the task and stage of interaction.

\paragraph{Stable preferences appeared more clearly in how users felt about elicitation than in whether they acted on it.}
We trained the estimator on each participant's first three sessions and tested it on their responses to EUIs in the later sessions. A participant's own history did not predict whether they acted on a mark much better than the pooled history across participants (Brier scores: 0.25, 0.16, and 0.26 with personal history vs.\ 0.28, 0.17, and 0.24 with pooled history for P2, P8, and P3, respectively). Personal history was more informative for how participants felt about the marks. For P2, personal history correctly predicted all later ratings as \say{glad} (error 0.00 vs.\ 0.90 with pooled history) and captured the negative direction of P8's ratings, while pooled history predicted them as slightly positive (own $-0.50$, pooled $+0.10$, observed $-0.40$).

\subsubsection{RQ2. How do users interpret and want to control a model of their elicitation preferences?}
\paragraph{Participants recognized their profiles.}
All three participants generally agreed with the profiles learned from their behavior. P8 said, \say{I get the broad strokes of what you're saying here and I do agree with them}, while P3 described his preference as \say{more questions asked in the beginning, like the plan, the lo-fi.} Some participants also noticed the adaptation before seeing their profiles. P8 felt a change around day 3 and thought the system was \say{responding to some of the feedback I was giving}, while P2 described only \say{a feeling that things were going well, though it's hard to explain}.

Participants' explanations also revealed aspects of their preferences that interaction behavior alone did not capture. P8 wanted more questions because \say{I didn't know whether or not it was actually looking to the internet.} P2 similarly valued seeing what the system was asking because \say{opening it is good in itself \ldots it feels reassuring.} Although she acted on only about half of the marks she saw, she still valued having them available. For both participants, wanting more elicitation was partly about understanding and checking what the system was doing, rather than simply wanting more opportunities to respond.


\paragraph{Participants preferred to steer their elicitation profile rather than individual techniques.}

When shown the six profile coordinates as sliders, all three participants preferred controlling these dimensions over the per-mark checklist they were exposed to in the last session of study 1. P8 described choosing among individual marks as \say{extra effort}, whereas an \say{easy slider} let him state a preference about himself, such as \say{I'll be more inquisitive.} P2 similarly felt that the axes were a more \say{human-feeling} way to describe her preferences.

All three participants also saw the profile as a cross-task baseline rather than something that should be relearned for every task. P8 said, \say{even though the tasks varied, I still had a preference for the type of things I wanted to interact with. I definitely want to maintain one singular profile.} P3 similarly said that it \say{generally should not be that different, may adjust a little for a task}, while P2 would \say{adjust somewhere between my own position and the average, never all the way to the other side.} Participants still wanted to tune the profile for particular tasks. For example, P8 wanted more elicitation for research and personal-data tasks, but less when he wanted the system to make decisions for him. Rather than separate profiles for each task, participants preferred one profile they could adjust when needed. Together, these responses suggest a model in which a user-level profile provides a default, while the task can still shift how elicitation is generated.

%% file: sections/08_discussion.tex
\section{Discussion} 

Our goal was to conceptualize how to design and generate Elicitive User Interfaces to inform how future GenUI systems can better personalize interfaces to support the user. We discuss the broader implications that our investigation leads to and limitations to explore in future work.


\subsection{Capturing User Feedback through Generative User Interfaces}
A longstanding challenge in interactive systems is how to collect feedback that helps systems better understand their users. For GenUI, we have seen several approaches to gathering this feedback~\cite{kim2026maru, shaikh2025gum, lam2026jit, jelly}. EUIs create a new opportunity to collect feedback by tying a user's response to the task, the current interface, the decision being elicited, and even the alternatives that were presented. In this sense, EUIs can provide more contextualized and interpretable data than isolated interaction traces. Over time, we believe that such data can be used not only to build models of how users prefer to be elicited, but also, in more controlled settings, to study UI preferences as they arise during the use of a GenUI and within the context of the user's task.

At the same time, our findings also show that these interaction signals are not always straightforward to interpret. For example, participants also valued instances of elicitation they did not act on, because they confirmed the current interface or prompted them to think about other possibilities. This points to an important role for HCI in GenUI learning systems in that beyond collecting more interaction data, we need to design feedback mechanisms that make the meaning and limits of those signals more legible and interpretable. We see EUIs as one way to support this for GenUI systems, and future work can further explore how such feedback should be captured.


\subsection{Designing to Generate More Than The Interface}



As we see a proliferation of generative interfaces hit consumer applications such as Claude and ChatGPT, the growing sentiment is that interfaces are cheaper and easier to generate.
However, we believe that generative interfaces also open new demands for designing interactions for GenUI.
In this paper, we found that GenUI does not invite opportunities for users to discover and convey what they want in the interface---a need that becomes more important when interfaces are more adaptive.
We are curious to see what new interaction design demands GenUI surfaces, and how we must develop methods around this activity of ``GenUI Design.''

One possible design method might involve: (1) \textit{identifying} a unique demand that generative interfaces raise that today's conventional interfaces do not, (2) \textit{designing} a design space of interfaces or interactions that GenUI generates, and finally (3) \textit{instructing} GenUI to intelligently select and compose the most relevant design based on the user's context.
Our work followed this approach to designing elicitation in generative interfaces. First, we realized that today's interfaces do not design interactions to elicit user needs, because users almost never modify today's interfaces due to the friction of implementing them. However, generative interfaces can change at any moment, giving us the opportunity to build the design space of elicitation techniques for GenUI (\S\ref{4_design_EUI}) and instruct GenUI to draw from it (\S\ref{5_gen_EUI}).

We believe that there are more opportunities for future designers and researchers to draw inspiration from our approach to design better generative interfaces.
One opportunity is to design and instruct generative interfaces to provide richer feedforward or previews of what GenUI systems will generate according to a prompted change~\cite{vermeulen2013feedforward, min2025feedforwardgenerativeaiopportunities}.
Our work on EUIs surfaced this need, where we found that our \textit{placement} axis visualized how broadly a change affects across the interface, but did not illustrate clearly enough what the resulting UI would be.
Another opportunity is to explore how to make GenUI changes more transparent across repeated use.
Rather than only replacing UI elements, GenUI systems might instead dynamically generate visual traces within the interface that users can inspect and manipulate~\cite{hill1992editwearreadwear, scentedwidgets}.
We anticipate more factors in generating GenUI differently than today's conventional interfaces, and we hope that industry designers and HCI researchers can trailblaze these opportunities toward a rich practice of GenUI Design.

\subsection{What is the Future of UI/UX Design for Generative User Interfaces?}

We can generate user interfaces that support complete user workflows, follow UI design patterns, and build under design systems \cite{mobbin2026}.
Meanwhile, we are challenging our views of what a ``user interface'' is more than ever, with computer use agents fully automating non-trivial tasks such as editing videos and building 3D models \cite{openai2026gpt6astra} and world interface models generating interactive applications entirely through images and videos \cite{runway2026solaris, rivard2026neuralossimulatingoperatingsystems}.
In this turbulent time for user interface design, what will be the role of UI/UX Designers for GenUI?

Our learnings from EUIs lead us to believe that, just as developers have moved from writing code to instructing how coding agents build software, perhaps designers must also move from creating individual UI mockups to \textit{instructing how GenUI systems design interfaces}. 
For instance, among the participants we observed, power users and UI designers leveraged \sysname{}'s EUI sliders to personalize how it elicited what they wanted.
We suspect that steering GenUI in this way---by refining the design space it draws from---could help scope the boundaries of the types of interfaces GenUI systems generate.
Producing such instructions would let UI designers offer valuable insight to the developers engineering GenUI systems.
We see this as an opportunity for future work to explore the shifting role of UI Designers.

\subsection{Limitations and Future Work}

\subsubsection{Scope and Extensibility of Elicitation Design}



Our design space primarily characterizes how an unresolved decision should be shown to the user, such as how salient the elicitation is, when and how often it appears, how strongly it asks for a response, and where it is placed. \sysname{} is designed to determine what decisions are worth eliciting, but this process is largely handled by the generation model and heuristics in our pipeline. Thus, a well-designed elicitation can still ask about a less important decision to the user, while an important user need might not be surfaced. Participants reflected this distinction themselves, at times wanting more relevant questions, not just more questions. Future work could combine EUIs with context- and user-modeling approaches~\cite{shaikh2025gum, lam2026jit, yen2024memolet}. Such models could help determine what requires elicitation from the user, while our EUI design space determines how to ask for it. 

At the same time, there can be many more techniques and considerations that should go into designing an elicitation than our design space, and other techniques may reveal dimensions that we did not capture. Yet the axis-based structure of it makes extensions additive, where future work can identify new and independent dimensions and incorporate them as additional parameters in the generation pipeline. 

\subsubsection{Comparing EUIs with Other Generative Interfaces}
Our studies were designed to understand how users experience different forms of elicitation. We therefore did not directly compare \sysname{} against a GenUI with no elicitation or one that uses a single fixed elicitation technique. Thus, while our findings show that users respond differently to different elicitation designs, we cannot fully isolate the additional benefits that came from adaptively selecting. Future controlled studies could compare no elicitation, fixed elicitation, and adaptive elicitation to better clarify when the specific form of elicitation matters and when adapting it to the user, task, or interface provides additional benefit. Nevertheless, our findings show that how GenUI systems ask for input is itself an important design choice.

\subsubsection{Measuring Interface Effectiveness}

Our studies primarily evaluate whether EUIs help users surface and act on their preferences, and whether the elicitation itself is experienced as useful or welcome. These measures capture the effectiveness of elicitation, but they do not establish whether the resulting interface is ultimately more effective for the user’s task. A change that better matches a user’s immediate preference may not necessarily improve task performance, usability, decision quality, or longer-term satisfaction. Future work could therefore evaluate the full downstream effect of elicitation, from the preferences EUIs surface to the modifications users make and, ultimately, how those changes affect the quality and use of the resulting GenUI.

%% file: sections/09_conclusion.tex
\section{Conclusion}

In this research, we explored Elicitive User Interfaces as a way to design how users shape generative interfaces.
Our design space and studies show that EUIs can surface preferences, invite feedback, and reveal how users want to be involved in shaping what is generated. 
As GenUI turns interfaces from fixed artifacts to more adaptable ones, GenUI design must expand beyond generating the interface but also to designing interactions and feedback through which users can steer it. 
We hope EUIs provide a step towards a richer practice of GenUI design, shedding light into how HCI can pave the path for designing valuable interactions in an increasingly AI-mediated world.


%




%% file: sections/99_appendix.tex
\captionsetup{width=.8\textwidth}
\clearpage
\onecolumn

\section*{Appendix}
\renewcommand{\thesection}{A.\arabic{section}}
\renewcommand{\thesubsection}{\thesection.\arabic{subsection}}
\renewcommand{\thefigure}{A.\arabic{figure}}
\renewcommand{\thetable}{A.\arabic{table}}
\setcounter{section}{0}
\setcounter{figure}{0}
\setcounter{table}{0}
\setcounter{page}{1}

\section{Formative Study Details}
\label{app:formative-stimuli}
\subsection{Task 1 --- Learning: How credit scores work}
\begin{enumerate}
  \item ``I keep hearing about credit scores but don't really get how
        they work. Can you explain it?'' \\
        \emph{Generated:} explainer cards / factor breakdown
  \item ``So what actually makes the number go up or down?'' \\
        \emph{Generated:} checklist of score-affecting events
  \item ``If I wanted to improve mine, what would move it the most?'' \\
        \emph{Generated:} sliders + projection graph, score gauge
  \item ``What counts as a good vs.\ bad score anyway? What are the
        ranges?'' \\
        \emph{Generated:} score-range table / banded meter
\end{enumerate}

\subsection{Task 2 --- Structured writing: Scholarship cover letter}
\begin{enumerate}
  \item ``I need to write a cover letter for a scholarship application.
        Help me draft it.'' \\
        \emph{Generated:} fill-in template / form fields
  \item ``Here's my background and the scholarship: [background and
        criteria block]'' \\
        \emph{Generated:} drafted letter with editable text blocks
  \item ``Which parts of this actually make me stand out? Highlight
        what's strongest and what's filler.'' \\
        \emph{Generated:} annotated draft with strength highlighting,
        ranked list of points
\end{enumerate}

\subsection{Task 3 --- Planning: Birthday gathering under constraints}
\begin{enumerate}
  \item ``I'm hosting a small birthday gathering at my place in three
        weeks --- around 8--10 friends, casual. Help me plan it.'' \\
        \emph{Generated:} tabs + task checklist + prep timeline
  \item ``There'll be a couple vegetarians and I've got about \$150 ---
        can you work within that?'' \\
        \emph{Generated:} cost breakdown cards, menu list
  \item ``Can you show me how the \$150 breaks down across
        categories?'' \\
        \emph{Generated:} budget chart
\end{enumerate}



\clearpage
\section{\sysname{} Details}
\label{app:Ditto}
The implementation of \sysname{} and the design space skill files are included in the supplementary material. 

\begin{figure}[h]
    \centering
    \includegraphics[width=\linewidth]{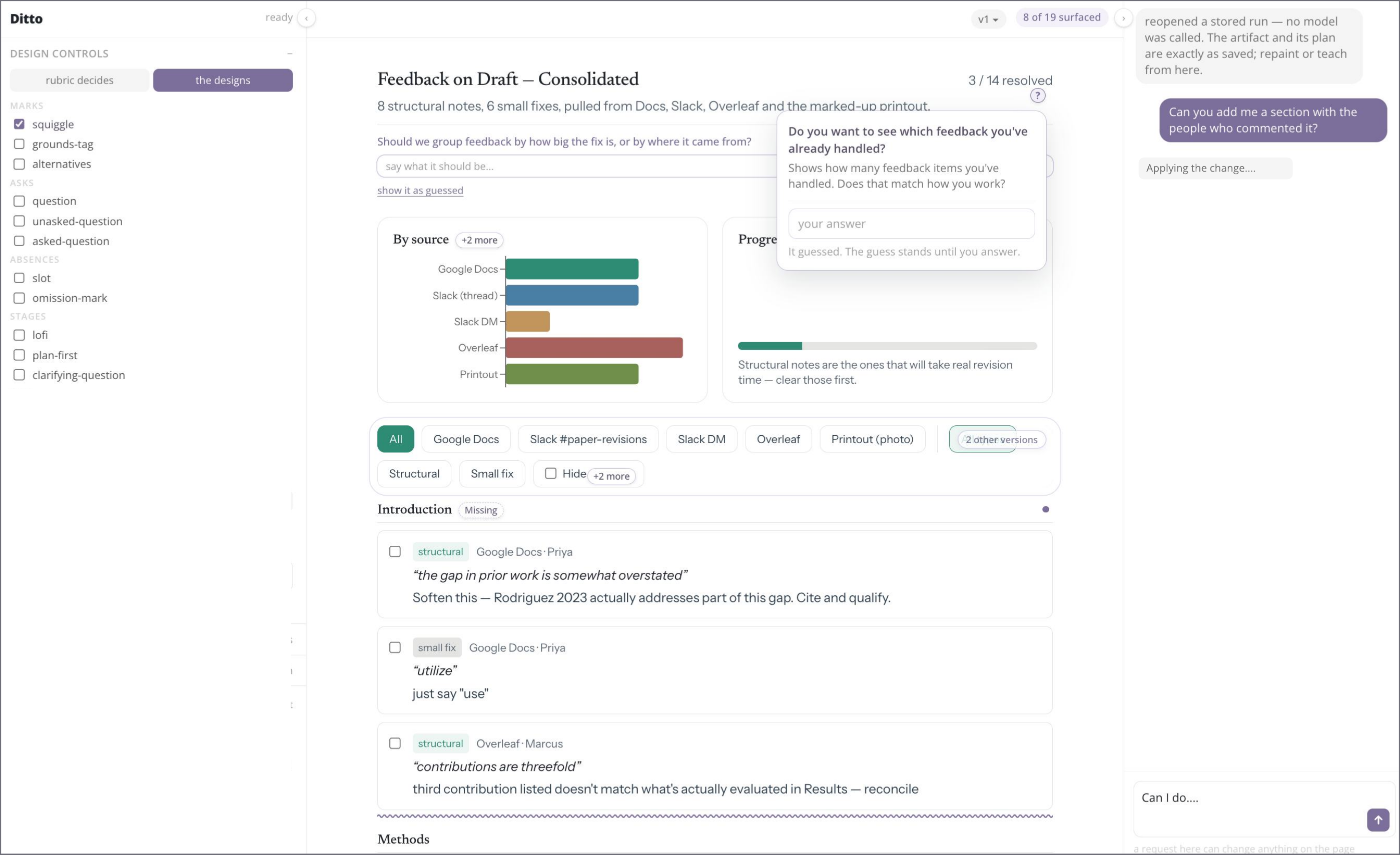}
    \caption{The \sysname{} probe interface.}
    \label{fig:placeholder}
\end{figure}

\clearpage
\section{User Study Details}
\subsection{User Information}
\begin{table}[H]
  \caption{Study 1 participants. Education
  is formal education (major, minor, or coursework) in design or frontend
  development. The last two columns are self-reported experience designing
  UIs in design tools (e.g.\ Figma) and implementing UIs in code. Participants marked $^\dagger$ returned for Study 2.}
  \Description{A table with one row per participant, P1 through P12, and
  seven columns: gender, age, position, LLM usage frequency, formal
  education in design or frontend development, experience designing UIs in
  design tools, and experience implementing UIs in code. Seven participants
  are men and five are women, aged 19 to 32. Eight use LLMs daily and four
  use them three to five times a week. Eight have formal education in
  design, frontend development, or both (P3, P4, P5, P6, P7, P10, P11,
  P12), and four have neither (P1, P2, P8, P9). Tool and code experience
  ranges from never (for example P8 and P9) to frequently (for example P3,
  P4, and P5). P2, P3, P4, and P8 are marked as having returned for
  Study 2.}
  \Description{A table summarizing the demographics and prior experience of
twelve Study 1 participants, P1 through P12. The table reports gender, age,
position, LLM usage frequency, formal education in design or frontend
development, experience with UI design tools, and experience implementing UIs
in code. Participants range from 19 to 32 years old and include seven men and
five women. Eight report using LLMs daily, and four report using them three to
five times per week. Eight have formal education in design, frontend
development, or both. Experience with design tools and UI implementation ranges
from never to frequently. P2, P4, and P8 are marked as participants who
returned for Study 2.}
  \label{appendix:participants}
  \begin{tabularx}{\linewidth}{@{}llllXlll@{}}
    \toprule
    & \textbf{Gender} & \textbf{Age} & \textbf{Position}
    & \textbf{LLM use} & \textbf{Education} & \textbf{Design tools} & \textbf{UI in code} \\
    \midrule
    P1            & W & 24 & Master's student   & Daily        & None            & Never       & Frequently \\
    P2$^\dagger$  & W & 23 & Master's student   & Daily        & None            & A few times & A few times \\
    P3$^\dagger$  & M & 29 & Master's student   & Daily        & Design + FE     & Frequently  & Frequently \\
    P4  & W & 26 & PhD student        & Daily        & Design + FE     & Frequently  & Frequently \\
    P5            & M & 23 & Recent graduate    & Daily        & Design + FE     & Frequently  & Frequently \\
    P6            & M & 26 & Master's student   & 3--5$\times$/wk & Frontend     & A few times & Frequently \\
    P7            & M & 23 & Undergraduate      & Daily        & Frontend        & A few times & Frequently \\
    P8$^\dagger$  & M & 24 & Undergraduate      & 3--5$\times$/wk & None         & Never       & Never \\
    P9            & M & 19 & Undergraduate      & 3--5$\times$/wk & None         & Never       & Never \\
    P10           & W & 32 & PhD student        & 3--5$\times$/wk & Design (UI/UX) & A few times & A few times \\
    P11           & W & 29 & Graduate student   & Daily        & Design (UI/UX)  & Frequently  & A few times \\
    P12           & M & 23 & Undergraduate      & Daily        & Frontend        & A few times & Frequently \\
    \bottomrule
  \end{tabularx}
\end{table}

\subsection{Questionnaire and Interview Questions}
\begin{table}[H]
  \caption{The Study 1 questionnaire. Per-elicitation items were asked once
  per elicitation encountered, at the end of each session. A ``didn't
  notice'' option disabled the ratings for that elicitation and logged it as
  unseen. Session items were asked once per session, and the final items
  once, after each participant's last session.}
  \Description{A table listing the questionnaire items in three groups. The
  first group was asked once per elicitation encountered, at the end of
  each session: a one-to-five usefulness rating from not useful to very
  useful; a one-to-five rating of how it felt, from annoying to glad it was
  there; and the question of whether the idea was already on the
  participant's mind or seeing the elicitation gave them the idea. The
  second group was asked once per session: a one-to-five rating of how
  close the final page came to what the participant wanted; a more, keep as
  is, or drop choice for each elicitation technique encountered; and a
  free-text question asking whether there was anything they wanted
  different but never told the system, with four optional reasons. The
  third group was asked once, after the last session: whether they would
  use something like this for real work; whether the pages felt mostly the
  system's, a mix, or mostly theirs; whether working this way changed what
  they asked for; and pick lists for techniques they would want more of or
  would drop.}
  \label{tab:questionnaire}
  \begin{tabular}{@{}p{0.17\textwidth}p{0.42\textwidth}p{0.33\textwidth}@{}}
    \toprule
    & \textbf{Item} & \textbf{Response format} \\
    \midrule
    \multirow{3}{*}{\shortstack[l]{Per elicitation\\(each session)}}
    & Useful? & 1--5, ``not useful'' to ``very useful'' \\
    & How it felt & 1--5, ``annoying'' to ``glad it was there'' \\
    & Was this already on your mind, or did seeing it give you the idea?
      & ``already on my mind'' / ``seeing it gave me the idea'' /
        ``didn't care either way'' \\
    \addlinespace
    \multirow{3}{*}{\shortstack[l]{Per session}}
    & The page, as it ended up\ldots
      & 1--5, ``not what I wanted'' to ``exactly what I wanted'';
        optional free-text why \\
    & Next time, for a task like this\ldots\ (per technique encountered)
      & ``more'' / ``keep as is'' / ``drop'' \\
    & Anything you wanted different, but never told it?
      & free text; optional reason: ``didn't know how to say it'' /
        ``not worth the effort'' / ``didn't think it could'' /
        ``it was fine as is'' \\
    \addlinespace
    \multirow{4}{*}{\shortstack[l]{Final\\(after last session)}}
    & Would you use something like this for your real work?
      & ``probably not'' / ``for some tasks'' / ``yes, regularly'';
        free-text ``for what?'' \\
    & The pages you ended with felt\ldots
      & ``mostly the system's'' / ``a mix'' / ``mostly mine'' \\
    & Did working this way change what you asked for, or what you wanted?
      & ``not really'' / ``a little'' / ``yes, clearly'';
        free-text ``how?'' \\
    & Pick any you'd want more of, and one you'd drop
      & pick lists over the techniques this participant encountered \\
    \bottomrule
  \end{tabular}
\end{table}

\subsection{Specific Tasks Done by Users in Study 1}
\begin{table}[H]
  \caption{What each participant asked for, by task type. Participants
  personalized every query themselves from a task template (``Teach me
  how\ldots'', ``Help me plan\ldots'', ``Help me write\ldots''). Asterisks mark each participant's free session, where they configured the EUIs themselves.}
  \Description{A table with one row per participant, P1 through P12, and
  three columns for the three task types: learning, planning, and writing.
  Each cell gives the query that participant wrote for that task type, such
  as learning laundry techniques, planning a scholarship application, or
  writing a statement of purpose. One cell per row is marked as the free
  session, in which the participant chose the task type, except P10, whose
  three sessions were all assigned.}
  \Description{A table listing the Study 1 queries written by participants P1
through P12 for three task types: learning, planning, and writing. Each row
shows the participant's personalized query for each task type. Learning tasks
include topics such as laundry techniques, social computing, Korean, medical
image processing, and probability theory. Planning tasks include scholarship
applications, coursework, travel, events, and other personal plans. Writing
tasks include research summaries, statements of purpose, letters, essays,
proposals, scripts, and other written artifacts. An asterisk marks the session
in which the participant chose the task type themselves rather than receiving
an assigned task type.}
  \label{tab:queries}
  \begin{tabular}{@{}l>{\raggedright\arraybackslash}p{0.28\textwidth}>{\raggedright\arraybackslash}p{0.28\textwidth}>{\raggedright\arraybackslash}p{0.28\textwidth}@{}}
    \toprule
    & \textbf{Learning} & \textbf{Planning} & \textbf{Writing} \\
    \midrule
    P1  & laundry techniques for different fabrics, step by step
        & time management for a scholarship application
        & a summary of their research to appeal to a potential advisor$^*$ \\
    P2  & the history of social computing$^*$
        & master's coursework at the [institution]
        & a statement of purpose for a graduate lab application \\
    P3  & learning intermediate Korean, with the learning curve
        & a graduation plan for the master's program
        & a wedding congratulation letter for a friend abroad \\
    P4  & how Ozempic works, and the alternatives
        & preparing for a first pet corn snake$^*$
        & an interview plan for a factory sensor-dashboard project \\
    P5  & Roman empire history and memorable wars
        & an evening couple date$^*$
        & a way to track personal expenses \\
    P6  & preparing dishes from different countries
        & a trip to Brazil, with flights and hotels
        & a startup proposal to present to investors$^*$ \\
    P7  & PyTorch for deep learning, from CNNs to VAEs$^*$
        & a family meetup dinner: groceries, schedule, music
        & an essay on an artwork from the campus museum \\
    P8  & starting a large-scale farm business
        & a movie shoot for a short student film
        & a screenplay for a sci-fi horror anthology episode$^*$ \\
    P9  & gustation and olfaction in the nervous system$^*$
        & college courses across a double major, 21 credits
        & a 500-word sociolinguistics essay \\
    P10 & getting started with medical image processing
        & a three-day trip to Xiamen
        & a paragraph on the concept of the responsibility gap \\
    P11 & a personal researcher web page, based on their CV
        & a month in New Zealand in December
        & a script for a ten-minute conference talk$^*$ \\
    P12 & the key concepts of probability theory
        & a trip to Japan$^*$
        & an essay on whether the Fed should raise rates \\
    \bottomrule
  \end{tabular}
\end{table}

\subsection{Specific Tasks Done by Users in Study 2}

\begin{table}[H]
  \caption{What each Study 2 participant asked for, session by session. All
  tasks were the participants' own.}
  
  \Description{A table with one row per session, one through seven, and three
  columns for the returning participants P2, P3, and P8. Each cell gives
  the query that participant wrote for that session, such as a study
  interface, a travel plan, or a letter to a friend. A horizontal rule
  after session three marks where the preference profile was switched on.
  P2 ran six sessions, P3 seven, and P8 six; cells for sessions a
  participant did not run are empty.}
  
  \label{tab:study2-queries}

  \small

  \begin{tabularx}{\columnwidth}{
    @{}l
    >{\raggedright\arraybackslash}X
    >{\raggedright\arraybackslash}X
    >{\raggedright\arraybackslash}X
    @{}
  }
    \toprule
    & \textbf{P2} & \textbf{P3} & \textbf{P8} \\
    \midrule

    S1
      & an interactive study interface for the four Gospels
      & a diet plan toward an ideal weight by June
      & a one-month TOPIK study plan around class hours \\

    S2
      & comparing how NLP and HCI papers argue, section by section
      & a Chuseok holiday plan around [city]
      & a worksheet teaching students to write an investor pitch \\

    S3
      & tracing the evolution of a theory their research builds on
      & learning to build agentic AI with their Claude subscription
      & a worst-case projection of a US bond market collapse \\

    \midrule

    S4
      & a homecoming event for a student club, browsing venues
      & a plan for learning pastry cooking
      & an interactive infographic article website \\

    S5
      & a pop quiz generated from ML lecture slides
      & re-learning Japanese, years after passing N3
      & a pitch for a kids' TV show to start as a YouTube series \\

    S6
      & putting their music taste into words, starting from a canvas of songs
      & a long letter to a close friend back home
      & a career plan toward biotech work in the US \\

    S7
      &
      & a website for catching up on Pok\'emon game updates
      & \\

    \bottomrule
  \end{tabularx}
\end{table}